\documentstyle[twoside,fleqn,espcrc2,epsfig,rotating]{article}
\newcommand{\AmS}{{\protect\the\textfont2
  A\kern-.1667em\lower.5ex\hbox{M}\kern-.125emS}}

\title{$3$--Dimensional Approach to Hot Electroweak Matter
for $M_{Higgs} \leq 70$ GeV}

\author{M.~G\"urtler
\address{Institut f\"ur Theoretische Physik, Universit\"at Leipzig, Germany}%
, E.-M.~Ilgenfritz
\address{Institut f\"ur Physik, Humboldt-Universit\"at zu Berlin, Germany}%
        \thanks{Talk given by E.-M. Ilgenfritz}%
	\thanks{Supported by
		the DFG under grant  Mu932/1-2}%
, J.~Kripfganz
\address{Institut f\"ur Theoretische Physik, Universit\"at Heidelberg,
        Germany}%
        \thanks{Supported by
		the DFG under grant  We1056/2-3}%
, H.~Perlt$^{\ \rm a}$%
        \thanks{Supported by
		the DFG under grant  Schi422/2-3}
        and
A. Schiller$^{\ \rm a}$}

\begin{document}

\begin{abstract}
We study the electroweak phase transition by
lattice simulations of an
effective 3-dimensional theory, for a Higgs mass
of about $70$ GeV. Exploiting a variant of the
equal weight criterion of phase equilibrium,
we obtain transition temperature, latent heat
and surface tension and compare with $M_H \approx 35$ GeV.
For the symmetric phase, bound state masses and the
static force are determined and compared  with results
for pure $SU(2)$ theory.
\end{abstract}


\maketitle

\section{Introduction}

There is an old belief that the electroweak standard
theory possesses a first order phase transition
\cite{KirzhnitsLi}-\cite{Kajantieoct95} at some temperature of the order
of the $W$ mass.
This phase transition has become subject of intensive studies in
the last years, in particular its dependence on the mass of the
so far elusive Higgs boson.
One motivation was the phenomenological interest
in baryon asymmetry generation at the electroweak scale.
The transition has to be strong enough, both in order to accomplish
a sufficient rate of baryon generation during the transition and
to prevent the wash--out of baryon number after it is completed.
The present quantitative understanding of possible mechanisms
as well as experimental lower bounds for the Higgs mass make this
unlikely within the minimal standard model.

A second reason
was the wish to control the behaviour of perturbative
calculations of the effective action. This quantity is the appropriate tool
of (non--lattice) thermal quantum field theory
for dealing with symmetry breaking.
Infrared problems prevent a perturbative evaluation of
the free energy in the symmetric phase to higher loop order.
However, the true, non--perturbative
nature of
the symmetric phase will be characterized by massive
$W$-- and Higgs bound states instead of massless $W$ gauge bosons.
A selfconsistent
approach to provide masses across the transition, e.g. by gap equations,
can improve the ability to calculate
perturbatively the symmetric phase.
Gauge field condensates are another property of the symmetric phase,
expected to lower its
free energy density.

An independent method is needed to characterize
the electroweak phase transition
even within a pure  $SU(2)$ gauge--Higgs version
of the theory. This model has become a testfield to control
the validity of perturbative
predictions over a broad range of Higgs masses.
At present, one is interested to see whether the first order
transition ends somewhere around a Higgs mass $M_H \approx 100$ GeV.
Lattice simulations
 \cite{BunkEA}-\cite{Kajantieoct95} are not only able to describe
both phases starting from first principles but, moreover, make it possible to
put both phases into coexistence near   the phase equilibrium.
Thus one is able to measure directly quantities like
latent heat, surface tension, condensates
etc. quantifying the strength of the transition.

One approach to lattice calculations of the electroweak transition
is based
on an effective $3$--dimensional Higgs model.
It is attractive phenomenologically because it circumvents
the problem of putting chiral fermions on the lattice.
Due to dimensional reduction, fermions as well as non--static bosonic
modes contribute to the effective action.
In contrast to QCD, dimensional reduction should
work for the electroweak theory
around  and above the transition temperature
because $g^2$ is small.
For the electroweak phase transition this approach has been
pioneered by Farakos et al. (see e.g. \cite{FarakosEA,Kajantieaug95}).
This program aims at exploring the accuracy of dimensional reduction
at various
Higgs masses by comparing
various parameters of the transition with those of $4$--dimensional
lattice and
perturbative approaches.
Perturbation theory is necessary to relate the $4$--dimensional continuum
theory to the parameters of the dimensionally reduced theory and,
finally, to the bare coupling parameters of the lattice action.
Dimensionally reduced versions retain the remnant of
the temporal gauge field $A_0$ (as an adjoint Higgs field) or not (as
in this work).

Recently  \cite{physlett}  we have presented results obtained with
the $3$-dimensional
lattice model on the phase transition for $M_H \approx 35$ GeV.
Here we present some results of numerical work
on the more realistic
$M_H \approx 70$ GeV and compare them with the case of smaller Higgs mass.
As expected the first order nature has become weaker but is still evident.
In this talk we put our main emphasis on
ways to characterize the phase
equilibrium at finite lattice size  in order to
obtain the infinite volume limit
of the transition parameters,
and on non--perturbative features of the symmetric phase.
 The extrapolation to the continuum limit will be dealt with in
a forthcoming publication  \cite{future}.

\section{The Model}

We study the $SU(2)$--Higgs system with one complex Higgs doublet of
variable modulus.
The gauge field is represented by the unitary $2 \times 2$ link
matrices $U_{x,\mu}$ and the Higgs fields are written as
$\Phi_x = \rho_x  V_x$
($\rho_x^2= {1 \over 2} Tr(\Phi_x^+\Phi_x)$ is the Higgs modulus squared, $V_x$
an element of the group $SU(2)$).
The lattice action is
\begin{eqnarray}
S  &=& \beta_G \sum_p (1 - {1 \over 2} Tr U_p ) - \nonumber \\
 & &    \beta_H \sum_l
       {1\over 2} Tr (\Phi_x^+ U_{x, \mu} \Phi_{x + \mu}) + \nonumber \\
  & &  \sum_x  \big( \rho_x^2 + \beta_R (\rho_x^2-1)^2 \big)
\label{eq:action}
\end{eqnarray}
with
 $\beta_G = {4 / (a g_3^2)}$.
 In three dimensions the
lattice Higgs self--coupling is
 $\beta_R=  (\lambda_3/  g_3^2) \
    (\beta_H^2  / \beta_G)$,
 $g_3^2$ and $\lambda_3$ denote the   3--$d$ continuum gauge and Higgs self
couplings which are  3--$d$ renormalization group invariants.
They are related to the corresponding four dimensional couplings via
 $g_3^2 = T (g^2 + O(g^4))$ and
$\lambda_3 = T ( \lambda + O(g^4) )$.

String operators
like
\begin{equation}
E(l) = \Phi^+_x U_{x,\hat{\mu}}
U_{x+\hat{\mu} , \hat{\mu}} ... U_{x+(l-1) \hat{\mu} , \hat{\mu}} \Phi_{x +
l\hat{\mu}} \
\label{eq:strings}
\end{equation}
(of extension $l$)
are used to form Higgs and $W$-operators.
Actually the  $3--d$ masses (inverse correlation lengths) have been
obtained from the connected correlators between separated
''time slice''
(in $2 + 1$ dimensions !)
sums of ''spatial'' string operators, which
project out the proper $SU(2)$ and spin content.
The overlap with the lowest mass states is improved choosing an
extension $l=4$ for correlators in
the $W$ and also partly in the Higgs channel
(there also
$\rho_x^2$ is used).
A static force is (formally) defined using Wilson loops $W(R,T)$ of
asymmetric extensions $2\le R \le T \le L/2 $ ($L^3$ is the lattice size).

The bare lattice coupling parameters
$\beta_H$ and $\beta_R$
can be translated into a physical
temperature by a formula which generalizes the tree level relation between
the Higgs mass and the bare lattice parameters
including perturbative corrections
(too complicated to be given here).
We consider here the
quartic coupling $\lambda_3/g_3^2 =  0.0957$. According to
the tree level based relation
 $ \lambda_3 / g_3^2  \approx  \lambda / g^2  \approx
 M_H^2/ ( 8 M_W^2)$
 this would correspond to the case
of $M_H = 70$ GeV.
 Corrections to this approximate relation depend on details of the
dimensional reduction (loop order, the adjoint
Higgs field $A_0$ integrated out or not). Taking these corrections
into account ($4$--dimensional renormalization effects are neglected)
this coupling ratio corresponds to $M_H = 71.8$ GeV for
$g^2=4/9$.
For comparison, the lower Higgs mass had been simulated with
$\lambda_3/g_3^2 = 0.0239$, corresponding to $M_H = 38.3$ GeV.

The lattice Higgs self coupling $\beta_R$ is not fixed
but runs with $\beta_H$.
This is important for the multihistogram technique to be used.
In general, binning has to be performed
in the {\it two relevant parts} of the action
(corresponding to $\beta_H$ and $\beta_R$) and
some other observable.
Our data come from typically $100000$ configurations per
set of couplings (separated by one heat bath step mixed with
eight Higgs field   reflections).
Correlation measurements were separated by $10$ iterations.

\section{Localization of the phase transition}

In order to characterize the order of the phase transition at
Higgs mass $M_H \approx 70$ GeV
and to obtain the infinite volume critical hopping parameter
$\beta_{Hc}^\infty$
we have simulated lattice sizes $30^3$, $48^3$
and $64^3$. To control
the approach to the continuum limit we have chosen gauge
couplings $\beta_G=12$ and
$\beta_G=16$ (not presented here).
Lattice observables to monitor the phase
transition are the Higgs modulus
squared $\rho^2$ and the link contribution to the action
$E(1)$, but other pure  gauge field quantities
like the average plaquette or the Polyakov line give
$2$--state signals, too.

\begin{figure}[thb]
\epsfig{file=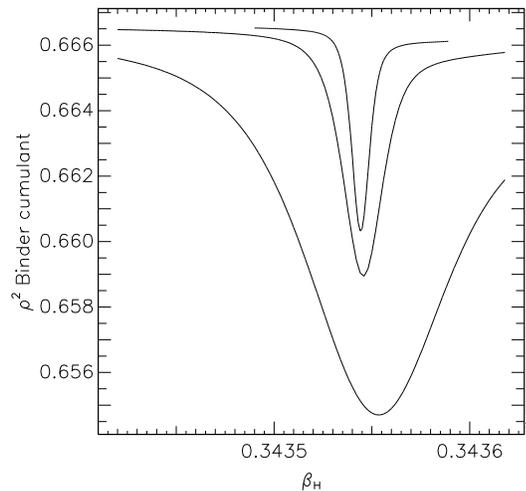,width=6.5cm,angle=90}
\caption{\sl Binder cumulants $B_{\rho^2}^L$ for $L=30$, $48$ and $64$
at $M_H \approx 70$ GeV}
\label{cumulant}
\end{figure}

We have used the Ferrenberg--Swendsen analysis and present
here only results obtained for histograms
of $\rho^2$. The corresponding Binder cumulants $B_{\rho^2}^L$ are shown
for the lattice sizes $L=30$, $48$ and $64$ in fig. \ref{cumulant}
The minima give the respective pseudocritical $\beta_{Hc}^L$. From
a linear extrapolation with $1/L^3$
we obtain the infinite volume limit $\beta_{Hc}^\infty = 0.3435433$.
For comparison,
the maxima of the $\rho^2$ susceptibility give
$\beta_{Hc}^\infty = 0.3435430$.
In the case of a $SU(2)$ Higgs model without $t$--quark
this value corresponds to
$T_c = 156$ GeV.
For comparison, we obtain $T_c = 98.7$ GeV for
$M_H \approx 35$ GeV.
Finiteness and shrinking of the Binder cumulant with increasing volume
are evidence for the first order nature of the transition at the larger
Higgs mass, too. The dip of the Binder cumulant reaches $0.6623$ in the
$V\to \infty$ limit clearly different from $2/3$.

Alternatively we have used and improved the equal weight
criterion for the localization of the transition.
In order to account for mixing of broken and symmetric
phases
at some floating $\beta_H$ {\it inside the metastability range}
we make the superposition ansatz for the histogram:
\begin{eqnarray}
P_m(\rho^2,\beta_H) &=& w_b P_b(\rho^2,\beta_H) +
w_s P_s(\rho^2,\beta_H) + \nonumber \\
& &w_{int} P_{int}(\rho^2,\beta_H).
\label{eq:histog}
\end{eqnarray}
For finite volumes there is also a contribution of
inhomogeneous states containing interfaces.
All histograms as well as the
sum of the appropriate
thermodynamic weights $w_i$ are normalized.
The histogram $P_m(\rho^2,\beta_H)$
is obtained by reweighting towards $\beta_H$,
merging all data by the Ferrenberg--Swendsen method.
The pure phase histograms should be
provided by the same technique from data taken
{\it outside
the metastable range}. Our data are scarce there, however.

\begin{figure}[thb]
\epsfig{file=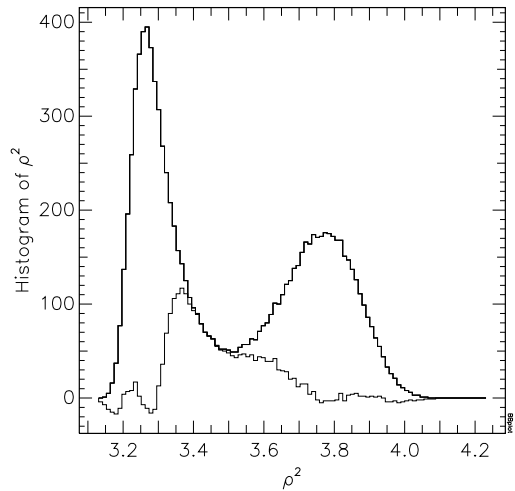,width=6.5cm,angle=90}
\caption{\sl Histograms $P_m$ and $P_{int}$ at $\beta_{Hc}$ for $L=64$
at $M_H \approx 70$ GeV}
\label{histogram}
\end{figure}

To obtain the pure phase histograms from runs in the metastability
region
one could use
a selection procedure removing tunneling and excursions to the
''wrong'' phase from the Monte Carlo history. This method has been used
to extract {\it pure phase correlation lengths} (see below) near   $T_c$.
It is unsuitable for the extraction of the weights $w_i$.
We have defined
{\it pure phase histograms} by Gaussian fits to the outer flanks of the
full
histogram $P_m(\rho^2,\beta_H)$, which allows to define the weights
as functions of $\beta_H$ and to determine the equilibrium point where
$w_s = w_b$.
More information is given by the probability of
mixed phase configurations, $w_{int}=1 -  w_b - w_s$ at $\beta_{Hc}$
and the slopes of $w_s$ and $w_b$ at intersection.
 The infinite volume
limit of the pseudocritical couplings is $\beta_{Hc}^\infty = 0.3435428$.
We show in fig.\ref{histogram} the full
histogram $P_m$ at $\beta_{Hc}^L$ for the $L=64$ lattice
and the contribution from inhomogeneous (mixed phase) configurations.
For a lattice of this size (in contrast to the smaller $L=48$ or $L=30$) the
pure phase histograms do not overlap, the gap is filled exclusively
by inhomogeneous states.

\section{Strength of the phase transition}

The present variant of the equal weight method for localizing the phase
transition has the advantage
to give an immediate estimate for the surface
tension. It is safer than Binder`s method which uses only the maxima
and the minimum of $P_m(\rho^2,\beta_{Hc})$. This is an important
improvement for very asymmetric (typical for Higgs transitions)
and strongly overlapping histograms (for all except the largest lattice sizes).
The excess free energy per unit area
of interphase surfaces
$\alpha$
can be obtained from
\begin{equation}
\frac{w_{int}}{w_b} = \exp(-\frac{A \alpha}{T_c}),
\label{eq:surftens}
\end{equation}
where the weights are taken at $\beta_{Hc}$ ($w_b = w_s$) and
$A$ is the surface between
the phases. The smallest surface $A$ is equal to $A = 2 L^2 a^2$.
Thus we obtain an estimate,
$\alpha /T_c^3 \approx 1.3 \times 10^{-4}$ at $L=64$.

The latent heat $\Delta\epsilon$ of the transition is given by the jump of
the average Higgs modulus squared according to
\begin{equation}
\frac{\Delta\epsilon}{T_c^4} =
\frac{M_H^2 }{8 T_c^2}\, g^2\, \beta_{Hc}\, \beta_G
\,\Delta\langle\rho^2\rangle.
\label{eq:latentheat}
\end{equation}
The phase separated histograms with respect to $\rho^2$ at $\beta_{Hc}$
give $\Delta\langle\rho^2\rangle$ slightly decreasing with
growing
lattice size.
The infinite volume extra\-polation with $1/L^2$ gives
$\Delta \rho^2_{\infty} = 0.499(10)$, which amounts to
$\Delta\epsilon/T_c^4 = 0.024(2)$. For comparison, we have obtained
$\Delta\epsilon/T_c^4 = 0.20(1)$ for the case $M_H =38.3$ GeV (at $\beta_G=12$,
too).

The equal weight method makes it possible to reconstruct the
free energy densities of the pure phases in the vicinity
of the phase equilibrium.
 The latent heat can then be expressed as the jump $\Delta \epsilon$ of
the energy density by
\begin{eqnarray}
\frac{\Delta\epsilon}{T_c^4} &=&
 \left(\frac{\beta_G g^2}{4}\right)^3\, T
\, \frac{d\beta_H}{dT} \times \nonumber \\
& &\frac{d}{d\beta_H}
\left(\frac{\log w_s}{L^3} -
      \frac{\log w_b}{L^3}\right)
\label{eq:wratio}
\end{eqnarray}
with derivatives taken at $\beta_{Hc}$.
Thus we obtain $\Delta\epsilon/T_c^4 = 0.0238$
for $L=64$.
Both methods give compatible
results for $\Delta \epsilon$
in the infinite volume limit within an $1/L^2$ extra\-polation.

The jump in $\langle \rho^2 \rangle$ at the critical temperature
(at $\beta_{Hc}$) may be translated into continuum units
\begin{equation}
\frac{\Delta \langle \phi^2 \rangle}{(g T_c)^2} =
\frac{\beta_{Hc} \, \beta_G}{4}\,
\Delta \langle \rho^2 \rangle .
\label{eq:condensat}
\end{equation}
This jump is independent of the $3$-dimensional renormalization scale
$\mu_3$, and is therefore more appropriate to consider than the
Higgs condensate itself. Extrapolating to infinite volume we find
$\sqrt{\Delta \phi^2} =0.706 \,g\,T_c$
 (at $\beta_G = 12$).
This value may be compared with predictions from $2$--loop perturbation theory
(in Landau gauge)
\cite{KripfganzEA,KripfganzEA2}
 $\sqrt{\Delta \phi^2}|_{pert} = 0.765 \,g\,T_c$.
One has to realize that $\sqrt{\Delta \phi^2}$ is not identical with
the condensate value $v(T_c)$. There is a relation
$\sqrt{\Delta \phi^2} = 0.904 \,v(T_c)$
at $\lambda_3/g_3^2 =  0.0957$
and with a renormalization scale $\mu_3 = v(T_c)$.

We conclude that
the phase transition at $M_H \approx 70$ GeV shows a clear
$2$--state signal, and quantities like the latent heat, the surface tension
and the Higgs condensate (all scaled by an appropriate power of $T_c$)
have a clearly non--vanishing infinite volume limit. For the latent heat
and the surface tension it must be
stressed that, contrary to the cases of very small Higgs mass, the
transition appears to be
weaker than predicted by perturbation theory.

\section{The strongly coupled symmetric phase}

This phase is characterized by a mass scale
$g^2 T$. It manifests itself e.g. in the string tension of the
dimensionally reduced
variant of the theory. As in the $4$--dimensional
theory
a force can be defined formally between static non--abelian charges through
Creutz ratios of Wilson loops.
In a previous work \cite{physlett} we have calculated
the force in the case of $M_H \approx 35$ GeV
in the symmetric as well as in the broken phase.
We have found a string tension
$\sigma = 0.11 \,(g^2T)^2$ in the symmetric phase,
whereas it vanishes in the broken
phase. This value has to be compared
to $\sigma = 0.13 \,(g^2T)^2$   for
the $3$--dimensional pure gauge theory \cite{Teper}.

\begin{figure}[thb]
\epsfig{file=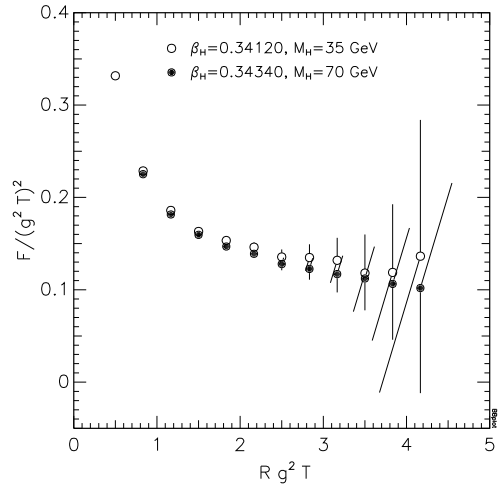,width=6.5cm,angle=0}
\caption{\sl Static force $F$ {\it vs.} distance $R$ for $M_H \approx 35$ GeV
and $M_H \approx 70$ GeV}
\label{force}
\end{figure}

In fig. \ref{force} we
show the force in the symmetric phase for the cases $M_H \approx 35$ GeV and
$M_H \approx 70$ GeV, both at $\beta_G=12$. From
these data for $M_H \approx 35$ GeV the above string tension
had to be fitted.
The errors are large but there is no evidence for
the expected
breakdown of confinement beyond some screening length.
For   $M_H \approx 70$ GeV
(due to the smaller $W$ mass in the symmetric phase compared to
the lighter Higgs case)
the perturbative contribution extends
further in distance.
Thus we can obtain only an upper bound for the
string tension, $\sigma \leq 0.1 (g^2 T)^2$.

There is another possibility to calculate the string tension
in the vicinity of the phase transition.
We consider the ratio of Wilson loops from {\it both sides of the phase
transition} which amounts to a
subtraction of the perturbative part.
This allows to anticipate the string
tension already at much smaller distance.
By this method we make more precise
for $M_H \approx 35$ GeV
the string tension $\sigma = 0.124 \,(g^2T)^2$
and predict
$\sigma = 0.078 \,(g^2T)^2$ for $M_H \approx 70$ GeV.

\begin{figure}[thb]
 \epsfig{file=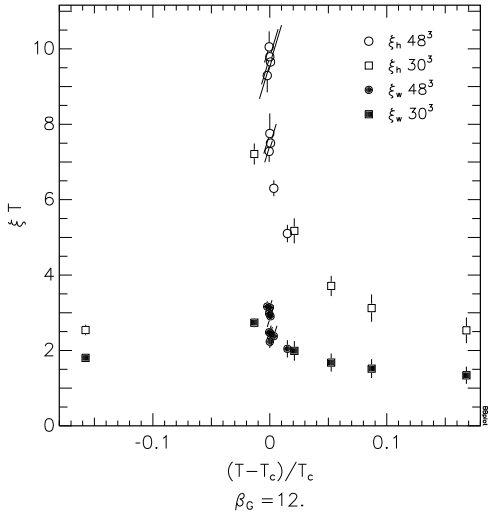,width=6.5cm,angle=0}
\caption{\sl Correlation lengths in Higgs and $W$ channel near   the
transition for
$M_H \approx 70$ GeV}
\label{corr}
\end{figure}

In order to characterize immediately the symmetric phase we have studied
masses in various channels. The behaviour of two correlation lengths
at $M_H \approx70$ GeV with
 Higgs and gauge boson quantum numbers  is presented
 in fig. \ref{corr}  in the vicinity of the phase transition.
 A discussion of these results together with the question of
gauge condensates are   postponed to Ref. \cite{future}.

 \par
{\bf Acknowledgements:} The calculations have been mainly
performed on the DFG--Quadrics QH2 parallel computer in Bielefeld.
 We wish to thank the system manager M.~Plagge for his help.
Additionally, we thank the council of HLRZ J\"ulich
 for providing CRAY-YMP resources.

\end{document}